\makeatletter
\newcommand{\dontusepackage}[2][]{%
  \@namedef{ver@#2.sty}{9999/12/31}%
  \@namedef{opt@#2.sty}{#1}}
\makeatother
\dontusepackage{subfigure}

\documentclass[]{article}

\usepackage{lmodern}
\usepackage{amssymb,amsmath}
\usepackage{ifxetex,ifluatex}
\usepackage[usenames,dvipsnames]{color}
\usepackage{fixltx2e} 
\ifnum 0\ifxetex 1\fi\ifluatex 1\fi=0 
  \usepackage[T1]{fontenc}
  \usepackage[utf8]{inputenc}
\else 
  \ifxetex
    \usepackage{mathspec}
    \usepackage{xltxtra,xunicode}
  \else
    \usepackage{fontspec}
  \fi
  \defaultfontfeatures{Mapping=tex-text,Scale=MatchLowercase}
  
\fi
\IfFileExists{upquote.sty}{\usepackage{upquote}}{}
\IfFileExists{microtype.sty}{%
\usepackage{microtype}
\UseMicrotypeSet[protrusion]{basicmath} 
}{}
\usepackage[margin=1.0in,bottom=1.5in]{geometry}
\usepackage[square,numbers,sort&compress]{natbib}
\usepackage{listings}
\usepackage{graphicx}
\makeatletter
\def\maxwidth{\ifdim\Gin@nat@width>\linewidth\linewidth\else\Gin@nat@width\fi}
\def\maxheight{\ifdim\Gin@nat@height>\textheight\textheight\else\Gin@nat@height\fi}
\makeatother
\setkeys{Gin}{width=\maxwidth,height=\maxheight,keepaspectratio}
\usepackage{caption}
\usepackage{float}
\usepackage{subfig}
\ifxetex
  \usepackage[setpagesize=false, 
              unicode=false, 
              xetex]{hyperref}
\else
  \usepackage[unicode=true]{hyperref}
\fi
\hypersetup{breaklinks=true,
            bookmarks=true,
            pdfauthor={},
            pdftitle={Parameterizing uncertainty by deep invertible networks, an application to reservoir characterization},
            colorlinks=true,
            citecolor=black,
            urlcolor=blue,
            linkcolor=black,
            pdfborder={0 0 0}}
\usepackage[preprint]{neurips_2019}
\usepackage{url}
\usepackage{booktabs}
\usepackage{amsfonts}
\usepackage{nicefrac}
\usepackage{microtype}

\usepackage{mathtools}
\newcommand{\bd}{\mathbf{d}}
\newcommand{\bu}{\mathbf{u}}
\newcommand{\bq}{\mathbf{q}}
\newcommand{\bz}{\mathbf{z}}
\newcommand{\F}{F}
\newcommand{\Z}{Z}
\newcommand{\M}{M}

\newcommand{\KL}{\mathrm{KL}}
\newcommand{\E}{\mathbb{E}}
\newcommand{\bm}{\mathbf{m}}
\newcommand{\bn}{\mathbf{n}}
\newcommand{\Normal}{\mathcal{N}}
\newcommand{\pdata}{p_{\mathrm{data}}}
\newcommand{\plat}{p_{\Z}}
\newcommand{\like}{p_{\mathrm{like}}}
\newcommand{\pnoise}{p_{\mathrm{noise}}}
\newcommand{\prior}{p_{\mathrm{prior}}}
\newcommand{\post}{p_{\mathrm{post}}}
\newcommand{\T}{T}
\newcommand{\norm}[1]{\left\lVert#1\right\rVert}

\title{Parameterizing uncertainty by deep invertible networks, an application
to reservoir characterization}
\author{Gabrio Rizzuti, Ali Siahkoohi, Philipp A. Witte, and Felix J.
Herrmann\\School of Computational Science and Engineering,\\Georgia
Institute of
Technology\\\texttt{\{rizzuti.gabrio,\phantom{\ }alisk,\phantom{\ }pwitte3,\phantom{\ }felix.herrmann\}@gatech.edu}}
\date{}

\begin{document}
\maketitle
\begin{abstract}
Uncertainty quantification for full-waveform inversion provides a
probabilistic characterization of the ill-conditioning of the problem,
comprising the sensitivity of the solution with respect to the starting
model and data noise. This analysis allows to assess the confidence in
the candidate solution and how it is reflected in the tasks that are
typically performed after imaging (e.g., stratigraphic segmentation
following reservoir characterization). Classically, uncertainty comes in
the form of a probability distribution formulated from Bayesian
principles, from which we seek to obtain samples. A popular solution
involves Monte Carlo sampling. Here, we propose instead an approach
characterized by training a deep network that ``pushes forward''
Gaussian random inputs into the model space (representing, for example,
density or velocity) as if they were sampled from the actual posterior
distribution. Such network is designed to solve a variational
optimization problem based on the Kullback-Leibler divergence between
the posterior and the network output distributions. This work is
fundamentally rooted in recent developments for invertible networks.
Special invertible architectures, besides being computational
advantageous with respect to traditional networks, do also enable
analytic computation of the output density function. Therefore, after
training, these networks can be readily used as a new prior for a
related inversion problem. This stands in stark contrast with
Monte-Carlo methods, which only produce samples. We validate these ideas
with an application to angle-versus-ray parameter analysis for reservoir
characterization.
\end{abstract}

\section{Introduction}\label{introduction}

A classical probabilistic setup for full-waveform inversion
\citep[FWI,][]{tarantola2005inverse} starts from the following
assumptions on the data likelihood:
\begin{equation}
    \bd=\F(\bm)+\bn.
\label{eq_like}
\end{equation}
 Here, $\bd$ represents seismic data, $\F$ is the forward modeling map,
$\bm$ collects the unknown medium parameters of interest, and $\bn$ is
random noise. We are assuming that $\bd$ and $\bn$ are independent
(knowing $\bm$), and $\bn\sim\pnoise(\bn)$ is normally distributed
according to $\Normal(0,I)$. Equation~\eqref{eq_like} then defines the
likelihood density function
\begin{equation*}
    \bd|\bm\sim\like(\bd|\bm)=\pnoise(\bd-\F(\bm)).
\end{equation*}
 In the Bayesian framework, the second ingredient is a prior
distribution on the model space
\begin{equation}
    \bm\sim\prior(\bm).
\label{eq_prior}
\end{equation}
 Since for seismic experiments we do not have direct access to the
Earth's interior (except for localized information in the form of well
logs), the prior is typically hand crafted based on, e.g., Thikhonov or
total variation regularization \citep[see, for
example,][]{esser2016tvr, peters2018pmf}. Alternatively, when this type
of information is available, a prior can be implicitly encoded via a
deep network \citep{mosser2018stochastic}.

The posterior distribution, given data $\bd$, is readily obtained from
the Bayes' rule:
\begin{equation}
    \post(\bm|\bd)=\like(\bd|\bm)\prior(\bm)/\pdata(\bd),
\label{eq_post}
\end{equation}
 where $\pdata(\bd)=\int\like(\bd|\bm)\prior(\bm)\,\mathrm{d}\bm$.
Sampling from the posterior probability~\eqref{eq_post} is the goal of
uncertainty quantification (UQ). Besides the usual maximum-a-posteriori
estimator (MAP)
\begin{equation}
    \bm_{\mathrm{MAP}}=\arg\max_{\bm}\post(\bm|\bd),
\label{eq_map}
\end{equation}
 being able to sample from the posterior allows to approximate the
conditional mean and point-wise standard deviation:
\begin{equation}
\begin{aligned}
    & \mu = \E(\bm|\bd)=\int\bm\,\post(\bm|\bd)\,\mathrm{d}\bm,\\
    & \sigma^2 = \E((\bm-\mu)^{.2}|\bd)=\int(\bm-\mu)^{.2}\post(\bm|\bd)\,\mathrm{d}\bm.
\end{aligned}
\label{eq_condest}
\end{equation}
 A great deal of research around UQ is devoted to Monte Carlo sampling
\citep{robert2004monte}, especially through Markov chains (MCMC) when
dealing with high-dimensional problems like FWI \citep[see,
e.g.,][]{malinverno2004, Bennett2005FastMU, malinverno2006, martin2012stochastic, sambridge2013transdimensional, Ely2018, marzuk2018, zhu2018, izzatullah2019bayesian, visser2019, zhao2019gradient}.
A Markov chain is a stochastic process that describes the evolution of
the model distribution, and is designed to match the target distribution
at steady state. A particularly egregious example is offered by the
so-called Langevin dynamics \citep{neal2011mcmc}:
\begin{equation}
    \bm_{n+1}=\bm_n-\frac{\varepsilon_n}{2}\nabla_{\bm}\log\post(\cdot|\bd)|_{\bm_n}+\eta_n,\quad\eta_n\sim\Normal(0,\varepsilon_nI),
\label{eq_langevin}
\end{equation}
 which is akin to stochastic gradient descent where the update direction
is perturbed by random noise $\eta_n$. Under some technical assumptions
on the step-length decaying \citep{welling2011bayesian}, collecting
$\bm_n$'s as in Equation~\eqref{eq_langevin} is equivalent to sample
from the posterior distribution (after an initial ``burn-in'' phase).
While gradient-based sampling methods such as Langevin dynamics are
gaining popularity in machine learning, we must face the computational
challenge given by the evaluation and differentiation of the seismic
modeling map involved in Equation~\eqref{eq_langevin}, which needs be
repeated for multiple sources thus resulting in an expensive scheme.
Nevertheless, source encoding and network reparameterization could
alleviate these issues
\citep{herrmann2019NIPSliwcuc, siahkoohi2020EAGEdlb, siahkoohi2020uncertainty}.

In this paper, we take an alternative route to UQ by employing
model-space valued deep networks defined on a latent space endowed with
an easy-to-sample distribution (e.g.~normal), as a way to mimic random
sampling from the posterior (when properly trained). This is made
possible by a special class of invertible networks for which the output
density function can be computed analytically
\citep{dinh2016density, kingma2018glow, lensink2019fully, kruse2019hint}.
As such, these networks encapsulate a new prior, which can be used for
subsequent tasks (as, for example, a similar imaging problem).

\section{Method}\label{method}

In this section, we discuss the theoretical foundations of the proposed
method for UQ via generative networks. We start by discussing a
well-known notion of discrepancy between probabilities, the
Kullback-Leibler divergence. Our program is suggested in
\citet{kruse2019hint}, and entails the minimization of the posterior
divergence with a known distribution pushed forward by an invertible
mapping. The resulting optimization problem is further restricted over a
special class of invertible networks, for which the log-determinant of
the Jacobian can be computed.

The Kullback-Leibler (KL) divergence of two probabilities defined on the
model space $\bm\in\M$, with density $p$ and $q$ respectively, is given
by
\begin{equation}
    \KL(p||q)=\int p(\bm)\log\dfrac{p(\bm)}{q(\bm)}\,\mathrm{d}\bm.
\label{eq_KL}
\end{equation}
 Let us fix, now, a normally distributed random variable
\begin{equation}
\bz\sim\plat(\bz),
\label{eq_Zlat}
\end{equation}
 taking values on a latent space $\Z$, and a deterministic map
\begin{equation}
\T:\Z\to\M.
\label{eq_T}
\end{equation}
 Selecting random samples $\bz$ and feeding it to $\T$ generates a
random variable $\T(\bz)$ with values in $\M$, which is referred to as
the push-forward of $\bz$ through $\T$. The push-forward density will be
denoted by
\begin{equation}
\bm\sim\T_{\#}\plat(\bm).
\label{eq_pushfw}
\end{equation}
 Our goal is the solution of the following minimization problem:
\begin{equation}
    \min_T\KL(\T_{\#}\plat||\post).
\label{eq_KLmin}
\end{equation}
 Note that the KL divergence is notoriously not symmetric, and the order
of the probabilities appearing in Equation~\eqref{eq_KLmin} is designed.
A more convenient form of Equation~\eqref{eq_KLmin} is
\begin{equation}
\begin{split}
    \KL(\T_{\#}\plat||\post)=\E_{\bz\sim\plat(\bz)}-\log\post(T(\bz)|\bd)\\-H(\T_{\#}\plat),
\end{split}
\label{eq_KLminEntr}
\end{equation}
 where
\begin{equation}
    H(p)=-\int p(\bm)\log p(\bm)\,\mathrm{d}\bm
\label{eq_entropy}
\end{equation}
 denotes the entropy of $p$. The loss function in
Equation~\eqref{eq_KLminEntr} ensures that the outputs of $\T$ adhere to
the data without producing mode collapse (e.g.
$T(\bz)\equiv\bm_{\mathrm{MAP}}$, a low-entropy configuration).

While the first term in the KL divergence can be approximated by sample
averaging, i.e.
$\E_{\bz\sim\plat(\bz)}f(\bz)\approx\frac{1}{N}\sum_{i=1}^Nf(\bz_i)$,
calculating the entropy of a push-forward distribution is not trivial
\citep[see, for instance,][]{kim2016deep}. In general, we typically can
only evaluate $T(\bz)$ without a direct way to explicitly compute its
density, as in Equation~\eqref{eq_pushfw}, for a given $\bm$. A
popular---but computationally expensive---workaround is based on
generative adversarial networks
\citep[GAN,][]{goodfellow2014generative}, where the discriminator is in
fact related to the generator push-forward density \citep{che2020gan}.
Other approaches involve variational autoencoders
\citep{kingma2013autoencoding} and expectation-maximization techniques
\citep{han2017alternating}. Nevertheless, if we focus on the specialized
class of invertible mappings $\T$ (whose existence requires that
$\dim\Z=\dim\M$ as manifolds), we obtain
\begin{equation}
    H(\T_{\#}\plat)=H(\plat)+\E_{\bz\sim\plat(\bz)}\log|\det J_{\T}(\bz)|
\label{eq_entropyinv}
\end{equation}
 thanks to the change-of-variable formula. Here, $J_{\T}(\bz)$ is the
Jacobian of $\T$ with respect to the input $\bz$.

Invertible networks are an enticing family of choice for $\T$. This is
especially true for sizable inverse problems, since they do not require
storing the input before every activation functions (such as ReLU) and
the memory overhead can be kept constant \citep{putzky2019invert}.
Furthermore, recent developments has lead to architectures that allows
the computation (and differentiation) of the log-determinants that
appears in Equation~\eqref{eq_entropyinv}. As notable examples, we
mention non-volume-preserving networks \citep[NVP,][]{dinh2016density},
generative flows with invertible 1-by-1 convolutions
\citep[Glow,][]{kingma2018glow}, and hierarchically invertible neural
transport \citep[HINT,][]{kruse2019hint}. These examples share the usage
of special bijective layers whose Jacobian has a triangular structure
\citep[Knothe-Rosenblatt transformations,][]{marzuk2018}. Hyperbolic
networks \citep{lensink2019fully} are another instance of invertible
maps, although volume preserving with null log-determinants. An
invertible version of classical residual networks (resnets) is also
offered in \citet{behrmann2018invertible}, which however requires an
implicit estimation of the log-determinant. Yet another example of
invertible architecture, though not allowing log-determinant
calculation, is discussed in \citet{putzky2019invert} for inverse
problem applications.

By restricting the problem in Equation~\eqref{eq_KLmin} to the family of
suitable invertible networks just described, we end up with the
following optimization:
\begin{equation}
\begin{split}
    \min_{\theta}\E_{\bz\sim\plat(\bz)}\frac{1}{2}\norm{\bd-\F(G_{\theta}(\bz))}^2-\log\prior(G_{\theta}(\bz))\\-\log|\det J_{G_{\theta}}(\bz)|,
\end{split}
\label{eq_uqprob}
\end{equation}
 where the unknowns $\theta$ represent the parameters (such as weights
and biases) of the invertible network
\begin{equation}
    G_{\theta}:\Z\to\M,\quad G_{\theta}^{-1}:\M\to\Z.
\label{eq_invnets}
\end{equation}
 We employ a stochastic gradient approach
\citep[ADAM,][]{kingma2014adam} with random selection of normally
distributed mini-batches of $\bz$'s. The expectation in
Equation~\eqref{eq_uqprob} is then replaced by Monte-Carlo averaging.

The solution of problem~\eqref{eq_uqprob} comes in the form of a network
$G_{\theta}$, whose evaluation over random $\bz$'s generates models
$\bm$'s as if they were sampled from the posterior distribution in
Equation~\eqref{eq_post}. After training, we can easily compute
conditional mean and point-wise standard deviation as in
Equation~\eqref{eq_condest}, or even higher-order statistics. Since
$G_{\theta}$ possesses a suitable architecture that explicitly encodes
its output density, we can reuse this result as a new prior for adiacent
regions, time-lapse imaging, and so on.

\section{Uncertainty quantification for reservoir
characterization}\label{uncertainty-quantification-for-reservoir-characterization}

We now apply the setup described in the previous section,
Equation~\eqref{eq_uqprob}, to full-waveform inversion for reservoir
characterization under some simplifying assumptions. We postulate a
purely acoustic model (constant density), with laterally homogeneous
properties (e.g., 1D). Waves propagate in a 2D medium. The Radon
transform yields the ``1.5-D'' wave equation
\begin{equation}
    (\bm(z)-p^2)\partial_{tt}\bu(t, z, p)-\partial_{zz}\bu(t, z, p)=\bq(t)\delta(z),
\label{eq_waveq}
\end{equation}
 where $\bu$ is a wavefield, $\bq$ a source wavelet, and $p$ the ray
parameter. Time-depth is indicated by $(t,z)$. Data is collected at the
surface $z=0$.

\begin{figure}
\centering
\includegraphics[width=0.500\hsize]{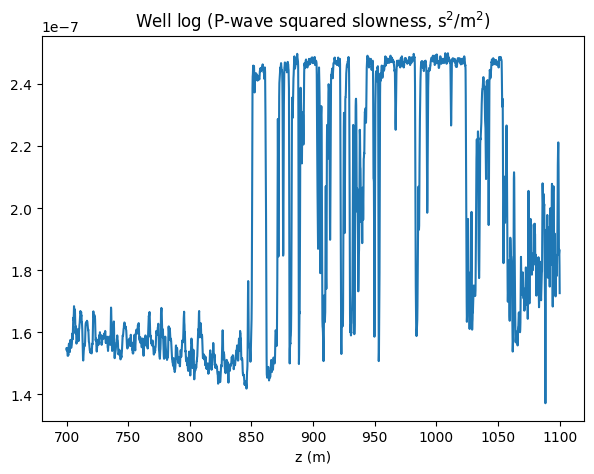}
\caption{Compressional velocity well log for the Sleipner synthetic
experiment}\label{log}
\end{figure}

\begin{figure}
\centering
\includegraphics[width=0.500\hsize]{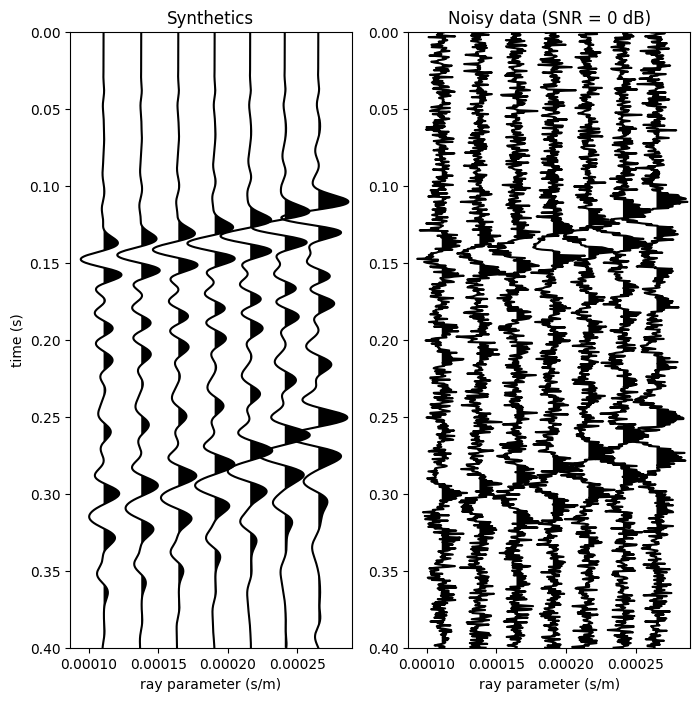}
\caption{Angle-versus-ray parameter data}\label{data}
\end{figure}

We synthetize a compressional velocity well-log from the open-source
Sleipner dataset in Figure~\ref{log} \citep[refer to][ for a previous
AVP study]{diSleipner}. See the acknowledgements for more info. The true
model is obtained by smoothing and subsampling the well-log slowness on
a $3$ m grid, and a background is obtained by severe smoothing. We
select a source wavelet with corner frequencies $(5, 30, 50, 80)$ Hz,
and a angle range from $\theta=12^{\circ}$ to $\theta=30^{\circ}$, where
$p=\sin\theta\sqrt{\bm(z=0)}$. Synthetic data are generated by
finite-differences, and random Gaussian noise is injected with
$\mathrm{SNR}=0$ dB. Data are shown in Figure~\ref{data}.

As of invertible architecture, we choose hyperbolic networks
\citep{lensink2019fully}, due to their inherent stability, augmented
with one activation normalization layer \citep[similarly
to][]{kingma2018glow} acting diagonally along the depth dimension. This
modification is necessary to obtain a non-volume preserving
transformation, that is with a non-null log-determinant (otherwise the
network entropy will remain constant, cf.~Equation~\ref{eq_entropyinv}).

\begin{figure}
\centering
\includegraphics[width=0.500\hsize]{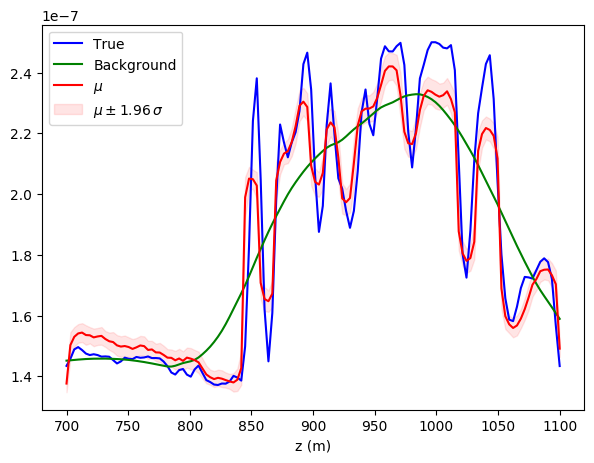}
\caption{Inversion results with confidence intervals}\label{minv_depth}
\end{figure}

\begin{figure}
\centering
\captionsetup[subfigure]{labelformat=empty}
\subfloat[(a)]{\includegraphics[width=0.500\hsize]{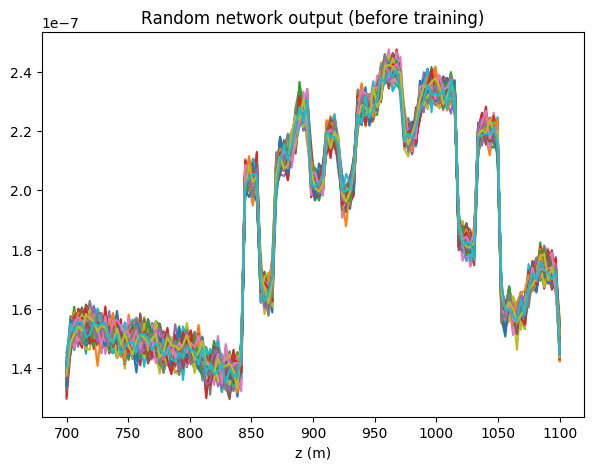}}
\subfloat[(b)]{\includegraphics[width=0.500\hsize]{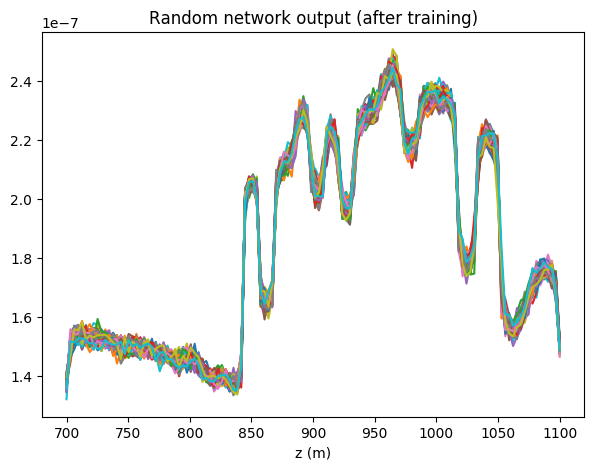}}
\caption{Network realizations: (a) before training (network initialized
as Gaussian perturbation of the MAP); (b) after
training}\label{minv_samples}
\end{figure}

\begin{figure}
\centering
\includegraphics[width=1.000\hsize]{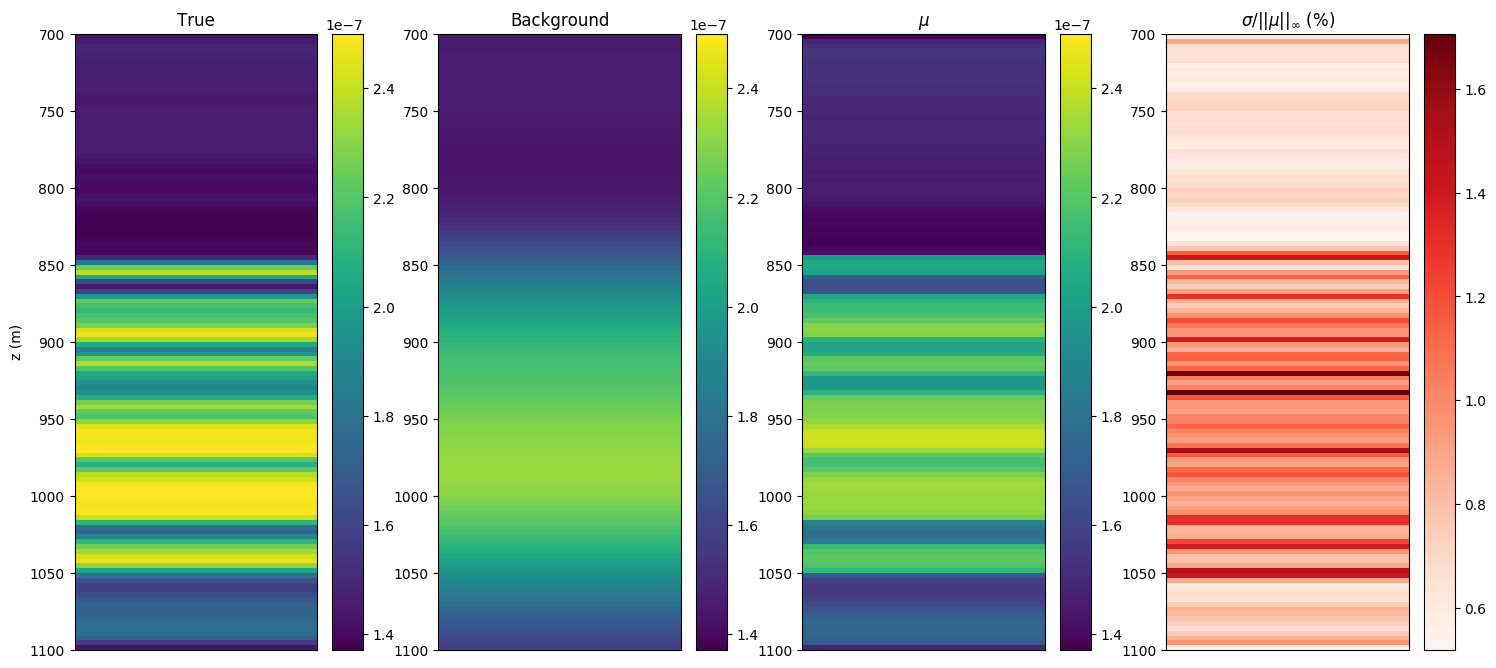}
\caption{Uncertainty quantification, structural comparison}\label{minv}
\end{figure}

The inversion results are compared in Figure~\ref{minv_depth}. We plot
the conditional mean $\mu$, as defined in Equation~\eqref{eq_condest},
with confidence intervals determined by the pointwise conditional
standard deviation $\sigma$. The conditional $\mu$ and $\sigma$ are
calculated with sample average of the network outputs $G_{\theta}(\bz)$.
Additionally we show multiple samples from the estimated posterior
before and after training (Figure~\ref{minv_samples}), which highlight
the sharpening of the distribution around the mean once optimized. Note
that $G_{\theta}$ was initialized as Gaussian perturbations of the MAP,
cf.~Figure~\ref{minv_samples}(a). A structural comparison between true
model and conditional statistics is made apparent in Figure~\ref{minv}.
As expected, the standard deviation has generally higher values where
the background model is further from the truth (the deeper portion), and
displays peaks at the reflector positions. We also observe that
uncertainties tend to grow with depth.

\section{Conclusions}\label{conclusions}

We presented an uncertainty quantification framework alternative to more
classical MCMC sampling, where an invertible neural network is
initialized according to a prior distribution and trained to approximate
samples from the posterior distribution, given some data. This can be
achieve by minimization of a probability discrepancy loss (the
Kullback-Leibler divergence) between the posterior and the network
output distribution. We discussed a full-waveform inversion application
for reservoir characterization, which demonstrates the ability of this
network to qualitatively capture the uncertainties of the problem.
Contrary to MCMC methods, the result can be reused as a new prior for
related inverse problems since, by construction, we are able compute the
density function of the output distribution. Despite the aim of this
project being faithful posterior sampling, we remark that the network
architecture do bias the uncertainty analysis \citep[sometimes favorably
as in deep prior regularization,][]{siahkoohi2020uncertainty}, and
should be carefully factored in. On the computational side, each
iteration of stochastic gradient descent requires the calculation of a
seismic gradient for the models of a given batch. Therefore, a feasible
extension to 2D would certainly need randomized simultaneous sources.
Future work will involve elastics, incorporation of hard constraints (as
opposed to priors used as a penalty term), and field data applications.

\section{Acknowledgments}\label{acknowledgments}

We are grateful to Equinor and partners for providing open access to the
Sleipner data through the CO$_2$ storage data consortium
(\href{https://co2datashare.org/}{co2datashare.org}).

\section{Related materials}\label{related-materials}

In order to facilitate the reproducibility of the results herein
discussed, a Julia implementation of this work is made available on the
SLIM github page:

\href{https://github.com/slimgroup/Software.SEG2020}{github.com/slimgroup/Software.SEG2020}

Several invertible network architectures are implemented as a Julia
package in:

\href{https://github.com/slimgroup/InvertibleNetworks.jl}{github.com/slimgroup/InvertibleNetworks.jl}

\bibliography{biblio_rizzuti2020SEGuqavp}

\end{document}